\documentclass[referee]{mn2e}

\usepackage{graphicx}
\usepackage{amsmath}

\title[On the origin of sub-TeV gamma-ray pulsed emission from rotating neutron stars]
{On the origin of sub-TeV gamma-ray pulsed emission from rotating neutron stars}
\author[W. Bednarek]
{W. Bednarek\\
Department of Astrophysics, The University of \L \'od\'z,
ul. Pomorska 149/153, 90-236 \L \'od\'z, Poland \\
bednar@astro.phys.uni.lodz.pl}
\begin{document}

\date{Accepted . Received ; in original form }

\pagerange{\pageref{firstpage}--\pageref{lastpage}} \pubyear{2007}

\maketitle

\label{firstpage}

\begin{abstract}
Intriguing sub-TeV tails in the pulsed $\gamma$-ray emission from the Crab pulsar have been recently discovered by the MAGIC and VERITAS Collaborations. They were not clearly predicted by any pulsar model.
It is at present argued that this emission is produced by electrons in the Inverse Compton process occurring either in the outer gap of the pulsar magnetosphere or in the pulsar wind region at some distance from the light cylinder. We analyse another scenario which is consistent with the basic features of this enigmatic emission. It is proposed that this emission is caused by electrons accelerated very close to the light cylinder where the
$e^\pm$ plasma can not saturate induced huge electric fields.
Electrons reach energies sufficient for production of hard $\gamma$-ray spectra in the curvature radiation process. Due to different curvature radii of the leading and trailing  magnetic field lines, the $\gamma$-ray spectra from separate pulses should extend to different maximum energies. The scenario can also explain the lower level $\gamma$-ray emission from the interpulse region (between P1 and P2)
observed in the Crab pulsar light curve.
Moreover, we argue that pulsars with parameters close to the Vela pulsar should also show pulsed emission with the cut-off at clearly lower energies ($\sim$50 GeV) than that observed in the case of the Crab pulsar. On the other hand, such tail emission is not expected in pulsars with parameters close to the Geminga pulsar. The model also predicts the tail $\gamma$-ray emission extending up to $\sim$50 GeV from some millisecond pulsars with extreme parameters such as PSR J0218+4243 and PSR J1823-3021A.
\end{abstract}

\section{Introduction}
The modern satellite telescopes (Fermi-LAT, AGILE) have detected over a hundred pulsars at energies between $\sim$100 MeV to $\sim 10$ GeV (e.g. Abdo et al.~2010a). These pulsars show a variety of light curves but their spectra are generally well described by a differential power law with an exponential cut-off at about a few GeV.
In fact, such behaviour has been predicted by the 
models considered for the pulsed $\gamma$-ray emission, i.e. the polar cap (Ruderman \& Sutherland~1975, Daugherty \& Harding~1982, Venter et al.~2009), the outer gap (Cheng et al.~1986, Romani~1996) or the slot gap (Arons~1983, Harding \& Muslinov~1998). 
However, observations of the Crab pulsar, performed with the MAGIC Cherenkov telescope, reported clear emission at energies $\sim$25 GeV (Aliu et al.~2008) and the evidence of the signal at even larger energies (above $\sim 60$ GeV, Albert et al.~2008). This emission component, extending to a few tens of GeV, has been also confirmed by the Fermi-LAT observations of the Crab pulsar (Atwood et al.~2009). Farther observations with the MAGIC and VERITAS telescopes have measured the pulsed $\gamma$-ray emission from the Crab extending up to $\sim 400$ GeV (Aliu et al.~2011, Aleksi\'c et al.~2011a), providing detailed spectral information concerning both $\gamma$-ray pulses (Aleksi\'c et al.~2011b). Whether this sub-TeV pulsed $\gamma$-ray emission connects naturally to the maximum at GeV energies or creates a separate spectral component is at present unclear.

These results attracted much attention since they were not commonly expected in the standard versions of the pulsar models.
More geometrically advanced versions of the outer gap model
postulate the appearance of the power law tails in the $\gamma$-ray spectrum of the Crab pulsar caused by the comptonization of the synchrotron UV-X radiation from the gap by secondary cascade $e^\pm$ pairs (e.g. Hirotani~2011, Aleksi\'c et al.~2011a, Lyutikov et al.~2011).
However, detailed modelling of the pulse emission extending up to at least $\sim 400$ GeV have not been provided up to now.  
It has been proposed that such sub-TeV pulsed $\gamma$-ray emission should also appear as a result of comptonization of the soft magnetospheric radiation by $e^\pm$ pairs already in the pulsar wind beyond the light cylinder radius (Bogovalov \& Aharonian~2000, Kirk et al.~2002, Aharonian et al.~2012). 
Lyutikov et al.~(2011) have discussed possible production of sub-TeV $\gamma$-ray tails by electrons accelerated within the outer gap region in the curvature radiation mechanism, concluding that the process is not efficient enough for production of $\gamma$-rays above $\sim$100 GeV
if the outer gap electric potentials are considered. 
However,they also suggested that if the acceleration process is
very efficient, electrons might produce sub-TeV $\gamma$-rays in the  curvature process. In this paper we investigate this suggestion in more detail considering a specific scenario for the origin of the hard $\gamma$-ray tail emission observed in the Crab pulsar.
We propose that electrons can reach energies large enough for production of sub-TeV curvature $\gamma$-rays in the region very close to the light cylinder radius, where the induced electric fields can not be saturated by the $e^\pm$ pair plasma. We show that such scenario is consistent with the main features of the pulsed $\gamma$-ray emission from the Crab pulsar.
Moreover, it gives clear predictions for possible appearance of such hard $\gamma$-ray emission in the case of other $\gamma$-ray pulsars. The basic features of this scenario can be tested by the future observations with the modern Cherenkov telescopes (HESS, MAGIC, VERITAS and also planned CTA).

\section{General scenario for the pulsed sub-TeV $\gamma$-ray emission}

We assume that the pulsed structure of the $\gamma$-ray emission from rotating neutron stars (pulsars) is due to the emission from a single magnetic pole. For an illustration, we show in Fig.~1 two critical magnetic field lines (marked by LL - leading line and TL - trailing line), extending to the light cylinder and anchored in a single magnetic pole. These lines bound in the opposite direction. The dipole structure of the magnetosphere of rotating neutron star can extend only to the light cylinder radius, $R_{\rm LC} = cP/2\pi = 4.78\times 10^7P_{10}$ cm, where $P = 10P_{\rm 10}$ ms is the period of rotation of the neutron star (NS) and $c$ is the velocity of light. Close to (and beyond) the light cylinder, the structure of the magnetic field has to gradually change to toroidal one as described by the asymptotic solution of the oblique split monopole (Bogovalov~1999).
The detailed structure of the transition region beyond the light cylinder is not well known. It is expected to be rather narrow with the characteristic thickness scaled with the light cylinder radius
(see Sect. 2.1 in Petri~2011). 
Therefore, in our scenario we also scale the curvature radius of the magnetic field lines, $R_{\rm c}$, at the light cylinder with the light cylinder radius. We introduce the scaling factor, $\eta$, which links $R_{\rm c}$ to $R_{\rm LC}$  according to $R_{\rm c} = \eta R_{\rm LC}$. 
Note that the leading and trailing magnetic field lines has to connect smoothly to the toroidal structure of the magnetic field in the wind region on this same distance scale in the transition region  beyond the light cylinder (as e.g. modelled by Petri~2011). Therefore, 
the curvature of the LL lines have to be smaller than the curvature of the TL lines (see Fig.~1).
Due to this difference the acceleration process of electrons along different lines is expected to be saturated at different maximum energies. We show that because of this difference in the curvature radii the energies of curvature photons produced by electrons accelerated in the region of the TL lines are expected to be significantly larger than those produced by electrons accelerated along the LL lines. Since the parameter $\eta$ can not be at present determined based on any theory, as an example we apply the values in the range $\eta = 0.5-2$, i.e.  smaller value for the LL lines and larger value for the TL lines. The comparison of the $\gamma$-ray spectra calculated in terms of such scenario with the future observations of the cut-offs in the $\gamma$-ray spectrum of separate pulses of the Crab pulsar should constrain the value of the parameter $\eta$, providing interesting constraints on the models of the pulsar magnetospheres close to the light cylinder.

The considered scenario predicts that the peaks in the pulsar light curves should have different spectra and structure.  We identify the emission from the LL lines with the pulse P1 in the Crab pulsar light curve. On the other hand, the pulse produced along the TL lines is responsible for the emission of the pulse P2 in the Crab pulsar light curve. Acceleration of electrons can also occur along the intermediate magnetic field lines (IL lines) which are closer to the magnetic pole. 
In this case the acceleration process of electrons is expected not to be so efficient since the acceleration gap along IL lines is likely to be narrow. This region of the gap is expected to be responsible for the $\gamma$-ray emission observed from the interpulse region of the Crab pulsar light curve.

In principle our scenario might be responsible not only for the hard tail $\gamma$-ray emission but also for the whole lower energy $\gamma$-ray emission observed from the Crab pulsar. In such a case, most of the  $\gamma$-ray emission is expected to be synchronized in phase. On the other hand, a lower energy part of the spectrum (at and below an exponential cut-off) could be produced in another mechanism (operating closer to the pulsar) which had been already mentioned in the  Introduction. In fact, some of the inner magnetosphere models (i.e. outer and slot gap models) predict $\gamma$-ray emission from the parts of the magnetosphere close to the critical magnetic field lines (LL and TL). These lines are directly linked to the emission region at the light cylinder considered in this scenario. Therefore, reasonable synchronization in phase, but not exact (see Abdo et al.~2010b, Aliu et al.~2011, Aleksi\'c et al.~2011), between emission coming from these two different regions is expected.

\begin{figure}
\vskip 8.truecm
\includegraphics{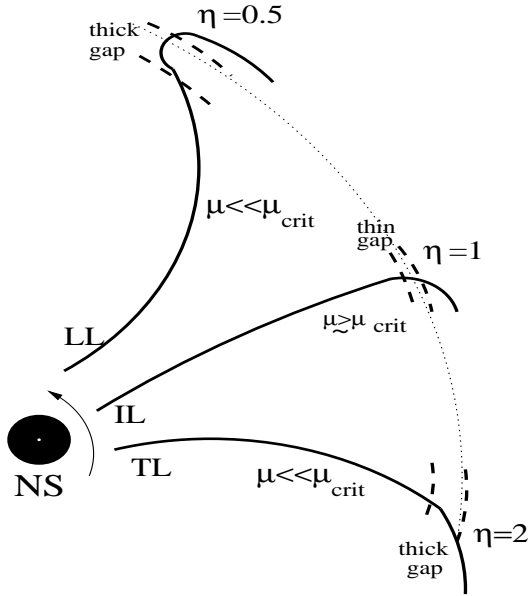}
\caption{Schematic representation of the pulsed $\gamma$-ray emission geometry in the case of a single magnetic pole model (not to scale). The magnetic field lines extend from the magnetic pole of the rotating neutron star (NS) up to the light cylinder radius (marked by the dotted curve) having a dipole structure. The direction of rotation of the NS is marked by an arrow. In such case, the leading magnetic field lines (marked by LL) have to change its geometry very significantly at the region of the light cylinder. On the other hand, the trailing magnetic field lines (TL) are stretched out. We also show the example intermediate line (marked by IL)
which curvature is expected to be intermediate between those of the LL and TL lines, e.g its radius can be close to the light cylinder radius ($\eta = 1$). 
We assume that electrons are accelerated efficiently close to the light cylinder radius within the so called {\it light cylinder gap}. The light cylinder gap is thick in the region of the LL and TL lines but it is thin in the region of the IL lines. The thickness of the gap is defined by the multiplicity of pair plasma arriving from the inner pulsar magnetosphere to the light cylinder radius. For LL and TL lines, the multiplicity is significantly below the critical multiplicity, $\mu << \mu_{\rm crit}$ (see for details Sect.~3.1). For the IL lines, the  multiplicity is $\mu\geq \mu_{\rm crit}$.  Electrons, accelerated in the gap, suffer curvature energy losses producing $\gamma$-rays.}
\label{fig1}
\end{figure}
\section{Pulsed $\gamma$-rays from curvature process}

We make use of the classical scenario for the pulsar magnetosphere defined in the work by Goldreich \& Julian~(1969, the Goldreich and Julian (GJ) model). In this model, the pulsar magnetosphere is filled with plasma which charge density is  
required to maintain co-rotation of the magnetic field lines.
In fact, this classical scenario is expected not be viable in the part of the magnetosphere which is charge separated (in the region between the critical and the last closed magnetic field lines, see Goldreich \& Julian~1969). However, it might be still valid in the rest of the open magnetosphere, i.e. along the magnetic field lines which are above the null surface (Spitkovsky~2011).
Therefore, we limit our considerations to the part of the magnetosphere between the magnetic pole and the critical line.
 
We assume that in the region close to the light cylinder a huge electric fields are generated in the magnetosphere rotating with the velocity close to the velocity of light. In a small region around $R_{\rm LC}$, these electric fields can not be saturated by the polarisation of the $e^\pm$ plasma since the required Goldreich and Julian plasma density should increase to infinity (see Goldreich \& Julian~1969). The electric field intensity close to the light cylinder can be of the order of,
\begin{eqnarray}
E_{\rm f} = ecB\approx 300B_{\rm LC}\approx 2.8\times 10^9B_{12}P_{10}^{-3} ~~~{\rm eV~cm^{-1}},
\label{eq1}
\end{eqnarray}
\noindent
where the magnetic field strength at the light cylinder is $B_{\rm LC}
= B_{\rm NS}(R_{\rm NS}/R_{\rm LC})^3$, $R_{\rm NS} = 10^6$ cm and $B_{\rm NS} = 10^{12}B_{12}$ G are the radius and the surface magnetic field strength of the NS, $e$ is the electron charge.

On the other hand, electrons lose energy on the curvature radiation,
\begin{eqnarray}
{\dot P}_{\rm c} = 2e^2\gamma^4/(3R_{\rm c}^2)\approx 4.2\times 10^{-23}\gamma^4/(\eta P_{\rm 10})^2~~~{\rm eV~cm^{-1}},
\label{eq2}
\end{eqnarray}
\noindent
where $\gamma$ is the Lorentz factor of electrons.
By comparing energy gains of electrons with their energy losses on the  curvature radiation ($E_{\rm f} = {\dot P}_{\rm c}$), we estimate the equilibrium Lorentz factor of electrons at the light cylinder gap,
\begin{eqnarray}
\gamma_{\rm eq} = 9\times 10^7 (B_{12}\eta^2/P_{10})^{1/4}.
\label{eq3} 
\end{eqnarray}
\noindent
Electrons can reach such Lorentz factors provided that the width of the light cylinder gap is large enough.
As we discuss below the gap along specific magnetic lines is constrained by the pair plasma production processes in the magnetosphere below the light cylinder radius.

\subsection{The thickness of the light cylinder gap}

The potential drop through the light cylinder gap has to be able to
accelerate electrons to the equilibrium energies ($E_{\rm eq} = m_{\rm e}\gamma_{\rm eq}$). As we mentioned above, the thickness of the gap
depends on the density of pairs which enter the light cylinder region
from the pulsar site. The calculations of the cascade processes in the inner magnetosphere of pulsars (below the light cylinder) show that the density of created pair plasma can overcome the Goldreich \& Julian density by a factor of the order of $\mu = n/(n_{\rm GJ, 0} = \Omega B/2\pi c)\sim 10^{(2-5)}$  (Harding \& Muslinov~2011, Hibschman \& Arons 2001, Ruderman \& Sutherland 1975).
These multiplicities are still about an order of magnitude lower than required in order to account for the observed radiation from e.g. the Crab Nebula (e.g. de Jager et al. 1996, Arons~2011).

We assume that electrons are accelerated in the electric field of the order of that given by Eq.~1 on the distance scale corresponding to the thickness of the gap, $\Delta R = R_{\rm LC} - R_{\rm b}$, where $R_{\rm b}$ is the distance of the bottom border of the acceleration gap from the NS. The density of plasma arriving to the light cylinder from the inner magnetosphere, defined by 
$\mu$, has to be lower than the Goldreich \& Julian density at the light cylinder region, described by the full formula $n_{\rm GJ} = n_{\rm GJ,0}/[1 - (R/R_{\rm LC})^2]$ (see Goldreich \& Julian~1969). 
Note that Goldreich \& Julian density diverges to infinity at the light cylinder radius. Thus, the acceleration gap has to appear even for very large multiplicity of pair plasma produced in the inner magnetosphere. Such scenario seems to be reasonable provided that the current induced by the dense plasma close to the light cylinder does not produce strong enough magnetic field which becomes comparable to the dipole magnetic field. Note that the GJ density formula has been derived based on the assumption of pure dipole structure of the magnetic field. It seems that this is still the case since in the GJ model the energy density in the plasma even at the light cylinder is smaller by a factor of $\sim$10$^4$ than energy density of the dipole magnetic field
(see e.g. review by Spitkovsky 2011).
 
To estimate the thickness of the light cylinder gap, we assume the validity of the formula for the Goldreich \& Julian density and compare it with the density of pair plasma
escaping from the inner magnetosphere (defined by coefficient $\mu$),
\begin{eqnarray}
\mu = [1 - (R_{\rm b}/R_{\rm LC})^2]^{-1}, 
\label{eq4}
\end{eqnarray}
\noindent
By inverting the above equation, we estimate the location of the bottom border of the gap and the thickness of the light cylinder gap,
\begin{eqnarray}
\Delta R = R_{\rm LC} - R_{\rm b} = R_{\rm LC}(1 - \sqrt{1 -1/\mu})\cong R_{\rm LC}/2\mu. 
\label{eq5}
\end{eqnarray}
We can estimate the critical value of the pair multiplicity for which the potential drop through the light cylinder gap is enough to accelerate electrons to the equilibrium energies. The following condition has to be met, $E_{\rm f}\cdot \Delta R\approx E_{\rm eq}$. 
The critical value of the pair multiplicity, derived from the above condition, is
\begin{eqnarray}
\mu_{\rm crit}\approx 1.5\times 10^3 P_{10}^{-7/4}B_{12}^{3/4}\eta^{-1/2}.
\label{eq6}
\end{eqnarray}
For pair multiplicities lower than the above value, electrons can reach the equilibrium Lorentz factors (given by Eq.~1) and produce high energy $\gamma$-rays in the curvature mechanism.
The pair plasma is not expected to be produced efficiently in the cascade processes in the inner pulsar magnetosphere along the last magnetic field lines, so called leading (LL) and trailing (TL) lines. Along these lines, the acceleration process of electrons and production of $\gamma$-rays are expected to occur already in the inner magnetosphere (e.g. the slot and outer gap models).
In these regions the multiplicity of created pairs should be rather
low in order to enable acceleration of electrons. On the other hand, the magnetic field lines closer to the axis of the magnetic pole (intermediate lines - IL) are expected to be more densely populated with the pair plasma (see e.g. the model by Ruderman \& Sutherland~1975). Therefore, we conclude that the light cylinder gap in the region of LL and TL lines is expected to be thick and accelerate electrons to the equilibrium Lorentz factors($\gamma_{\rm eq}$). On the other hand, the gap in the region of IL lines is expected to be significantly thinner. It can accelerate electrons marginally to $\gamma_{\rm eq}$ or below this value. Electrons in this part of the light cylinder gap are not able to extract energy equally efficiently as those accelerated in the 
gap along the LL and TL lines. Thus, they produce $\gamma$-ray on 
clearly lower level as observed in the interpulse region of the Crab pulsar light curve.

\subsection{Gamma-ray production} 

We assume that the magnetic field is significantly distorted from the dipole structure in order to met the toroidal structure in the pulsar wind zone. In such case, the curvature radius of the leading and trailing lines can differ from the light cylinder radius already in the light cylinder gap (see Fig.~1).
We can estimate the characteristic energies of curvature $\gamma$-rays produced by electrons which move with the equilibrium Lorentz factors through the gap (for the case of  $\mu < \mu_{\rm crit}$),
\begin{eqnarray}
E_\gamma^{\rm c} = 3hc\gamma_{\rm eq}^3/(4\pi R_{\rm c})\approx  
450 \eta^{1/2}B_{12}^{3/4}/P_{10}^{7/4}~~~{\rm GeV}.
\label{eq7} 
\end{eqnarray}
\noindent
The curvature radius of the leading lines is expected to be smaller than $R_{\rm LC}$ (i.e. $\eta < 1$) and the trailing lines larger than 
$R_{\rm LC}$ ($\eta > 1$). Therefore,
we identify the $\gamma$-ray emission from the 
pulse P1 in the Crab pulsar light curve as coming from the leading lines 
(lower maximum energies) and from the pulse P2 as coming from the trailing lines (larger maximum energies). 

The calculations of cascade processes in the inner pulsar magnetosphere envisage that the intermediate magnetic field lines are much densely populated by the pair plasma than the site lines along which gaps in the magnetosphere below the light cylinder radius can develop (e.g. Ruderman \& Sutherland~1975). 
The pair multiplicities in the region of the leading and trailing lines are expected to be significantly lower in order to allow for the gap formation along these lines in the magnetosphere below the light cylinder (see the slot gap and outer gap models). We assume that 
along the leading and trailing lines the pair multiplicities are below $\mu_{\rm crit}$. However, along the intermediate lines (between the leading and trailing lines) the pair multiplicities are much larger, i.e. they are close to (or even above) $\mu_{\rm crit}$. Therefore, 
electric potential through the light cylinder gap along the intermediate lines may not be enough to accelerate leptons to the
equilibrium Lorentz factors ($\gamma_{\rm eq}$).
As a result, $\gamma$-rays are expected to be produced less efficiently along the intermediate lines (the region between the leading and trailing lines) than along the leading and trailing lines.

The observations of the giant radio pulses, and also detailed calculations  of radiation processes in the inner magnetosphere (e.g. Timokhin 2010), argue for highly non-stationary production of pair plasma in the inner magnetosphere. Since the thickness of the gap along the intermediate lines is very sensitive on the multiplicity of pair plasma, it is expected that the acceleration of electrons in the light cylinder gap along the IL lines
should produce variable $\gamma$-ray emission in the Crab pulsar interpulse region. Therefore, $\gamma$-ray emission from the IL lines should be anticorrelated with the appearance of the giant radio pulses.
In contrust, the $\gamma$-ray emission from the pulse regions should not
show any features of correlation with the giant radio pulses.

\subsection{Other constraints on the radiation scenario}

The characteristic distance scale (the mean free path) on which electrons lose energy on curvature radiation is, 
\begin{eqnarray}
\lambda_{\rm c} = m_{\rm e}c^2\gamma_{\rm eq}/{\dot P}_{\rm c}(\gamma_{\rm eq})\approx 
1.7\times 10^4 \eta^{1/2} P_{10}^{11/4}/B_{12}^{3/4}~~~{\rm cm}.
\label{eq8}
\end{eqnarray}
\noindent
where $\gamma_{\rm eq}$ is given by Eq.~\ref{eq3} and ${\dot P}_{\rm c}$
is given by Eq.~\ref{eq2} applaying $\gamma_{\rm eq}$. 
Since $\lambda_{\rm c}$ is much shorter than the curvature radius of the 
magnetic field lines, $\gamma$-rays from curvature process can be emitted only within a cone defined by the angle,
\begin{eqnarray}
\Delta\alpha_{\rm cool}\approx \lambda_{\rm c}/\eta R_{\rm LC}\approx 3.6\times 10^{-4}P_{10}^{7/4}/(\eta^{1/2}B_{12}^{3/4})~~~{\rm rad.}
\label{eq9}
\end{eqnarray}
\noindent
On the other hand, the angular spread of this emission is also limitted by the thickness of the gap,
\begin{eqnarray}
\Delta\alpha_{\rm gap}\approx \Delta R/\eta R_{\rm LC}\approx 1/(2\mu\eta)~~~{\rm rad}
\label{eq10}
\end{eqnarray}
For the parameters of known pulsars and the multiplicity of pair plasma 
arriving from the inner magnetosphere, of the 
order of $10^2-10^5$, the angular spread of $\gamma$-ray emission is rather small, in agreement with the observations of the width of the pulses in the Crab pulsar light curve.
We expect that the width of pulses should rather correspond to the 
angular extend of the magnetic field lines along which the $\gamma$-ray pulsed emission is produced. These lines should physically link the light cylinder gap to the gaps appearing in the magnetosphere below the light cylinder. Therefore, similar location of the $\gamma$-ray pulses 
in the pulsar light curve is expected from the gaps below the light cylinder and those produced in the light cylinder gap.

The cooling process of electrons due to curvature radiation becomes inefficient if their energies drop below certain value $\gamma_{\rm min}$. This value can be estimated by the comparison of the mean free path for electrons on curvature radiation with the characteristic distance scale for the origin of this radiation which is the curvature radius of the magnetic field, i.e. $\lambda_{\rm c}\approx \eta R_{\rm LC}$. This condition is satisfied for
\begin{eqnarray}
\gamma_{\rm min}\approx 6.3\times 10^6\eta P_{10},
\label{eq11}
\end{eqnarray}
\noindent
Electrons with energies $\gamma_{\rm min}$ escape from the light cylinder region into the pulsar wind. These electrons can be farther cooled by Inverse Compton scattering of soft radiation as considered by Bogovalov \& Aharonian~(2000), Kirk et al.~(2002) and Aharonian et al.~(2012).

$\gamma$-rays produced at the light cylinder might have large enough energies to be absorbed in the local magnetic field close to the light cylinder ($\gamma B\rightarrow e^\pm$). Such absorption process should be important when the mean free path for $\gamma$-ray photons (see Eq.~3.3a in Erber~1966) becomes comparable to the light cylinder radius. This condition is equivalent to the approximate relation,
$(E_\gamma/2m_{\rm e}c^2)(B_{\rm LC}/B_{\rm cr})\approx 1/15.$
Note that the estimated number $1/15$ (see also Ruderman \& Sutherland~1975) does not differ significantly for the $\gamma$-ray pulsars in the range from the millisecond pulsars to the pulsars with the parameters (B,P) close to the Geminga pulsar. By inverting the above equation, we obtain the constraint on the surface magnetic field of the pulsars below which the curvature $\gamma$-rays can escape freely from the pulsar inner magnetosphere,
\begin{eqnarray}
B_{12}\approx 1.0P_{10}^{19/7}/\eta^{2/7}.
\label{eq12} 
\end{eqnarray}
\noindent
This condition is not very restrictive for the classical and millisecond pulsars.

By inverting Eq.~7, we obtain the constraints on the pulsar parameters for which curvature $\gamma$-rays can be produced with specific energies,
\begin{eqnarray}
B_{12}\approx 0.32P_{10}^{7/3}E_{100}^{4/3}/\eta^{2/3},
\label{eq13} 
\end{eqnarray}
\noindent
where $E_\gamma = 100E_{100}$ GeV is the energy of the $\gamma$-ray photon. This formula indicates that hard tails in the pulsed
$\gamma$-ray spectra at $\sim 100$ GeV can be only observed in the case of pulsars with rather extreme parameters such as observed in the Crab pulsar or some millisecond pulsars (for which $P\sim 1$ ms and $B\sim 10^9$ G), provided that the curvature radius of the magnetic field at the light cylinder gap is not very far from the light cylinder radius.

\section{Pulsed $\gamma$-rays from the Crab}

Up to now, hard $\gamma$-ray tails in the pulsed spectrum of only the Crab pulsar has been discovered (Aliu et al.~2008, Aliu et al.~2011, Aleksi\'c et al.~2011a,b). The emission from both pulses
(P1 and P2) differ significantly (Aleksi\'c et al.~2011b). $\gamma$-ray emission above $\sim$46 GeV is clearly weaker from the pulse P1 in comparison to the pulse P2. Moreover, there are some evidences of the extended tail in the light curve after the pulse P1(Fig.~1 in Aleksi\'c et. al.~2011b). 
The $\gamma$-ray spectrum of the pulse P2 extends up to $\sim$400 GeV without any evidences of the cut-off. In contrast, $\gamma$-ray emission from the pulse P1 is detected only to $\sim$200 GeV, with some evidences of the cut-off at higher energies. The width of the pulse P1 is estimated on $\Delta\alpha^{\rm Crab} = 0.025\pm 0.008$ and the pulse P2 on $0.053\pm 0.015$ (Aleksi\'c et al. 2011a).

We determine the basic parameters characterising our scenario by assuming the surface magnetic field of the Crab pulsar equal to $B_{\rm Crab} = 4\times 10^{12}$ G and its rotational period $P_{\rm Crab} = 33$ ms. For these parameters the equilibrium Lorentz factors of electrons accelerated in the light cylinder gap are $\gamma_{\rm eq}^{\rm Crab}\approx 9.4\times 10^7\eta^{1/2}$. The critical multiplicity of pair plasma arriving from the inner magnetosphere is 
$\mu_{\rm crit}^{\rm Crab}\approx 525\eta^{-1/2}$ and the characteristic energies
of $\gamma$-ray photons produced in the gap are $E_\gamma^{\rm Crab}\approx 160\eta^{1/2}$ GeV. The scenario underlined above requires that the multiplicity of pair plasma along the leading (LL) and trailing (TL) lines is clearly lower than $\mu_{\rm crit}^{\rm Crab}$. 
The widths of the pulses (P1 and P2), $\Delta\alpha^{\rm Crab}$, should be lower than estimated in Eq.~\ref{eq10}. This last condition, $\Delta\alpha^{\rm Crab} < \Delta\alpha_{\rm gap},$ allows us to constrain the lower limit on the multiplicity of pairs, $\mu > 1/2\eta\Delta\alpha^{\rm Crab}$, along the lines on which $\gamma$-rays are produced in the region of the pulses P1 and P2. This lower limit is
equal to $\mu > 20/\eta$ for the pulse P1 and $\mu > 10/\eta$ for the pulse P2. For the above estimated parameters, 
electrons are accelerated within the gaps along LL and TL lines to
the equilibrium Lorentz factors being in agreement with the angular extend of the $\gamma$-ray pulses from the Crab pulsar. Moreover, we conclude that the light cylinder gap along the intermediate (IL) lines has to be thin since the multiplicity of pairs along these lines is expected to be large (i.e. $\mu \geq \mu_{\rm crit}^{\rm Crab}$).  Therefore, the production of $\gamma$-rays along the IL lines in the region corresponding the interpulse region in the Crab pulsar light curve is expected to be less efficient in consistence with the observations.

In order to better envisage the potential of our scenario, and convince that the expected $\gamma$-ray emission features can be consistent with the observations of the Crab pulsar, we perform calculations of the curvature $\gamma$-ray spectra produced by electrons accelerated along the LL and TL lines, corresponding to the region of the pulse P1 and P2 in the Crab pulsar light curve. The pulsar electrodynamics close to the light cylinder is not precisely known. Therefore, the detailed calculations of the whole range of the $\gamma$-ray emission features are not possible on this stage. We consider a simple toy model
which at least can envisage the possibility of production of curvature $\gamma$-rays with energies observed in the pulse regions of the Crab pulsar.
We assume that electrons have monoenergetic energies (given by Eq.~\ref{eq3}) and the curvature of the magnetic field lines is described by the value of the parameter $\eta$ equal to 0.5 (for the LL lines) and equal to 2 (for the TL lines). The results are shown in Fig.~2. It is clear that for reasonable values of the curvature radii of the magnetic field lines, the curvature spectra can extend through the energy range observed by the MAGIC Collaboration from the Crab pulsar in the region of pulse P1 and P2 (Aleksic et al.~2012).

\begin{figure}
\vskip 5.truecm
\includegraphics{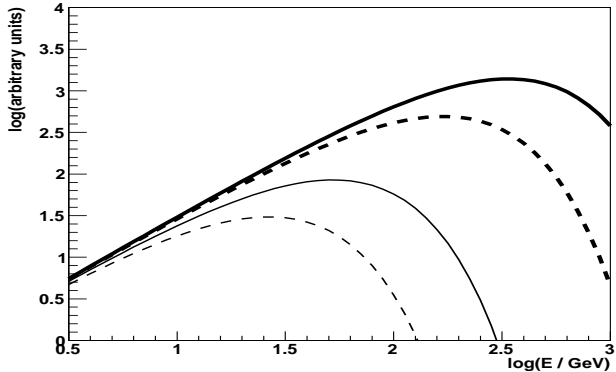}
\caption{The example $\gamma$-ray spectral energy distribution from the leading lines (dashed curves), corresponding to the  pulse P1 in the Crab pulsar light curve, and from the trailing lines (solid curves) corresponding to the pulse P2. The spectra are calculated applying a simple model in which monoenergetic electrons  with the equilibrium Lorentz factors (Eq.~3) produce curvature radiation in the magnetic field with the curvature radius defined by $\eta = 0.5$ (for the leading lines) and $\eta = 2$ (for the trailing lines). Two pulsars are considered: the Crab pulsar (thick curves) and the Vela pulsar (thin curves).}
\label{fig2}
\end{figure}
\section{Predictions for other pulsars}

Let us consider in a more detail the prospects for detection of
hard tails in the pulsed $\gamma$-ray spectrum of other pulsars.
According to our scenario, detection of the pulsed sub-TeV emission in the case of pulsars, with parameters less extreme than those of the Crab pulsar, is not so promissing. We consider two well known classical $\gamma$-ray pulsars, i.e. the Vela and Geminga pulsar (the strongest objects representing groups of pulsars detected in the GeV energies). The characteristic energies of $\gamma$-rays (Eq.~\ref{eq7}) estimated for the Vela pulsar ($B_{\rm NS} = 3.4\times 10^{12}$ G and $P = 89$ ms) and Geminga pulsar ($1.6\times 10^{12}$ G, $P = 237$ ms) are $E_\gamma^{\rm Vela}\approx 25\eta^{1/2}$ GeV and $E_\gamma^{\rm Geminga}\approx 0.3\eta^{1/2}$ GeV, respectively. For the curvature radii suitable for the Crab pulsar (e.g. $\eta = 0.5$ and 2), hard tail emission (above $\sim 30$ GeV) might be observed in the case of the Vela pulsar (see Fig.~2) but not in the case of Geminga pulsar. 
In fact, There are some evidences of the existence of such hard emission in the pulsed $\gamma$-ray spectrum of the Vela pulsar already in the Fermi-LAT measurements (see Fig.~7 in Abdo et al.~2010c and Lyutikov~2012).
Note also strong $\gamma$-ray emission from the interpulse region in the case of Vela pulsar light curve which suggest that pair multiplicity
along the intermediate lines is below or not far from the critical multiplicity also along some intermediate lines when the production of pair plasma in the inner magnetosphere is not efficient (periods between giant radio pulses). The hard $\gamma$-ray tails extending above $\sim 100$ GeV might only appear in the case of pulsars with less extreme parameters than those observed in the Crab pulsar provided that the magnetic field lines at the light cylinder are characterised by more extreme values of the curvature radii. 

Interestingly, the model predicts that some millisecond pulsars might 
also show hard $\gamma$-ray emission in the pulsed spectrum, similar to that observed in the Crab pulsar, since for them the characteristic energies of curvature photons are similar to that
expected for the Vela pulsar, i.e. 
$E_\gamma^{\rm c}\approx 140\eta^{1/2}B_9^{3/4}/P_1^{7/4}$ GeV, where
$B = 10^9B_9$ G and $P = 1P_1$ ms. 
For example, we consider in a more detail two MSPs: PSR J0218+4243 and
PSR J1823-3021A recently detected in the $\gamma$-rays by the Fermi-LAT telescope (Abdo et al.~2009, Abdo et al.~2011).
PSR J0218+4243 (surface magnetic field 4$\times 10^8$ G and the rotational period of 2.3 ms) is able to produce curvature $\gamma$-rays with energies $E_\gamma\approx 17\eta^{1/2}$ GeV and PSR J1823-3021A
(4.3$\times 10^{9}$ G and 5.44 ms) $E_\gamma\approx 22\eta^{1/2}$ GeV. Note that these values are similar to those derived for the Vela pulsar.
With the magnetic field curvature at the light cylinder described by the parameters applied for the Crab pulsar, 
the $\gamma$-ray emission from the trailing lines of  
these two pulsars might be detectable by the present MAGIC stereo and the future Cherenkov telescope systems (HESS II, CTA) with the threshold at $\sim 30$ GeV.

\section{Conclusions}

We propose that electrons accelerated close to the light cylinder 
(in the so called {\it light cylinder gap}) can reach sufficient energies for the production of curvature $\gamma$-rays with energies up to a few hundred GeV in the case of pulsars with parameters similar to the Crab pulsar. The scenario is able to explain the spectral differences between the Pulse P1 and P2 in the Crab pulsar light curve as a result of different curvature of the magnetic field lines at the light cylinder along the leading and trailing magnetic field lines.
The broad gaps close to the light cylinder appear as a result of efficient acceleration of particles close to the light cylinder where the density of the pair plasma cannot saturate the electric fields.
The model also predicts the appearance of lower level $\gamma$-ray emission in the interpulse region (betwen P1 and P2) as due to the acceleration of electrons in the gap which is narrower along the intermediate lines than along the leading and trailing lines.
The width of the gap along different lines is determined by the multiplicity of the pair plasma created within the inner pulsar magnetosphere along different magnetic field lines.
 
One of the interesting features of the proposed scenario is the relation of the efficient $\gamma$-ray emission to the magnetic field lines where the pair plasma density is low. Because of that, we predict that the hard $\gamma$-ray emission from the interpulse region cannot occur when the giant radio pulses are formed (caused by the non-stationary outflow of dense pair plasma through the pulsar magnetosphere).
In fact, the giant radio pulses are expected to be anticorrelated with 
the level of the hard $\gamma$-ray emission from the interpulse region.

The model also predicts hard pulsed $\gamma$-ray emission from pulsars with less extreme parameters than that of the Crab pulsar.
However, this emission is expected to extend to much lower energies.
For example, in the case of pulsars with the parameters close to the Vela pulsar, and also some millisecond pulsars (e.g. PSR J0218+4243 and PSR J1823-3021A), the pulsed $\gamma$-ray emission is expected to extend only to $\sim 50$ GeV. These hard $\gamma$-ray "tails" should appear only in the region of the trailing magnetic field lines. 
Our scenario does not predict any hard $\gamma$-ray emission in the case of pulsars with the parameters close to the Geminga pulsar.
Note that Inverse Compton pair cascade models for the sub-TeV $\gamma$-ray tail emission in the inner magnetosphere predict the cut-offs at much larger energies even in the case of the intermediate age pulsars since the primary cascade $\gamma$-rays are not expected to be absorbed so efficiently (e.g. Hirotani~2011, Du et al.~2012). This feature should allow to distinguish between these two models. 

In summary, our scenario predicts clear emission features for the pulsed $\gamma$-ray spectra from the classical pulsars with extreme parameters and from some millisecond pulsars. They are sometimes in contradiction to predictions of the popular Inverse Compton models for the origin of the sub-TeV pulsed $\gamma$-ray emission, in which $\gamma$-rays are produced by the secondary cascade $e^\pm$ pairs in terms of the outer gap scenario (Hirotani~2011, Lyutikov et al.~2011, Du et al.~2012) or by $e^\pm$ pairs outside the light cylinder radius in the pulsar wind (Bogovalov \& Aharonian~2000, Kirk et al.~2002, Aharonian et al.~2012). The future observations of the Crab pulsar (and also other pulsars) by the modern Cherenkov telescopes should allow us to distinguish between these different scenarios.

\section*{Acknowledgments}
I would like to thank the anonymous referee fro useful comments.
This work is supported by the grant through the Polish Narodowe Centrum Nauki No. 2011/01/B/ST9/00411.


\label{lastpage}
\end{document}